\documentclass[preprintnumbers,nofootinbib,aps,prl,floatfix,twocolumn]{revtex4}

\usepackage{epsfig,subfigure}

\usepackage{epstopdf}

\pdfoutput=1
\usepackage[utf8]{inputenc}
\usepackage{graphicx}
\usepackage{latexsym}
\usepackage{amsfonts}
\usepackage{amssymb}
\usepackage{amsmath}
\usepackage{slashed}
\usepackage{hyperref}
\usepackage{url}
\usepackage{color}
\usepackage{multirow,array}
\usepackage{float}
\usepackage{subfigure}

\newcommand{\vev}[1]{\langle {#1} \rangle}
\newcommand{\lsim}{\lesssim}
\newcommand{\gsim}{\gtrsim}

\newcommand{\eq}[1]{Eq.~(\ref{#1})}
\newcommand{\fig}[1]{Fig.~\ref{#1}}
\newcommand{\mP}{M_{\rm Pl}}

\newcommand{\ord}[1]{\mathcal{O}{(#1)}}
\newcommand{\beq}{\begin{equation}}
\newcommand{\eeq}{\end{equation}}
\newcommand{\bea}{\begin{eqnarray}}
\newcommand{\eea}{\end{eqnarray}}
\newcommand{\eps}{\varepsilon}

\newcommand{\nn}{\nonumber}

\newcommand{\uprime}{U(1)^{\prime}}

\newcommand{\maprime}{m_{A^{\prime}}}

\begin{document}

\date{\today}
\title{Getting a THUMP from a WIMP}

\author{Hooman Davoudiasl\footnote{email: hooman@bnl.gov}
}
\affiliation{Physics Department, Brookhaven National Laboratory, Upton, New York 11973, USA}

\author{Gopolang Mohlabeng\footnote{email: gmohlabeng@bnl.gov}
}
\affiliation{Physics Department, Brookhaven National Laboratory, Upton, New York 11973, USA}

\begin{abstract}
Producing an acceptable thermal relic abundance of dark matter with masses $\gg 10^2$~TeV is a challenge.  
We propose a novel mechanism where GeV-scale states establish a tiny thermal relic abundance for dark matter, which is later promoted to ultra massive status by a very light scalar. We refer to this dark matter as a THermal Ultra Massive Particle (THUMP). Direct detection of THUMPs can be naturally expected due to large scattering cross sections mediated by low mass states that couple THUMPs to the Standard Model.  Our model generically leads to signals for the associated GeV-scale states at accelerator experiments.

\end{abstract}

\maketitle
\section{Introduction\label{sec:intro}}

The search for clues to the identity of dark matter (DM) continues. So far, no experiment has uncovered conclusive evidence regarding the properties of DM and the range of possibilities 
for what it could be remains vast.  In this situation, one may be compelled to try and look for potential candidates wherever possible.  However, launching a search in every possible direction is not feasible and very often one needs some motivating factor to justify the effort.  Here, theoretical 
considerations could point to promising targets.  

The weak scale has been, for a long time, considered a likely place for new physics to emerge, in 
large part due to arguments based on a natural Higgs sector for the Standard Model (SM).  This scale is also typical for production of thermal relic DM, a fact that has strongly motivated searches for DM particles of corresponding mass.  However, after decades of dedicated searches for both 
new weak scale particles and DM at accelerators and direct detection experiments, no conclusive evidence of either has been uncovered.  While this may point to the elusiveness of such new states or 
the limitations of current experimental techniques, one could also start to entertain new 
or less examined possibilities.

Apart from the well-established weak scale, the scale of a possible quantum theory of 
gravity is another conceptually plausible place where new phenomena may arise.  This scale is taken to be set by the Planck mass $\mP \approx 
1.2 \times 10^{19}$~GeV, which is 
also typically assumed to mark the end of short distance physics.  Between the weak and Planck scales there could be several other motivated scales where new phenomena may appear.  For example, unification of known forces of Nature in a Grand Unified Theory
often points to masses $\gsim 10^{15}$~GeV for new states.  Also, the strong CP problem of QCD can be elegantly addressed through the Peccei-Quinn mechanism which typically points to scales 
$f_{PQ}\lsim 10^{12}$~GeV, to avoid conflict with cosmology (``overclosing'' the Universe) \cite{Preskill:1982cy, Abbott:1982af, Dine:1982ah}.  However, astrophysical constraints on the associated axion demand $f_{PQ}\gsim 10^9$~GeV \cite{Patrignani:2016xqp}.  
So, $10^9~\text{GeV}\lsim f_{PQ}\lsim 10^{12}~\text{GeV}$ provides another motivated window 
at very high scales (the axion could be a good DM candidate at the upper end of the window).

Although the above high scales provide motivation to consider DM masses well above the weak scale, 
DM relic abundance is generally assumed to be non-thermal for masses above $\sim 100$~TeV \cite{Griest:1989wd}.  In principle, one could construct such models of ultra massive DM, but generically the 
underlying physics and the signals of this DM may not be accessible, due to the high scales involved \cite{Chung:2001cb, Harigaya:2016vda, Harigaya:2016nlg, Kolb:2017jvz}. 
Different mechanisms for setting the thermal relic abundance of ultra massive DM, leading to various astrophysical signatures have been considered in the literature, see for example \cite{Berlin:2016vnh, Berlin:2016gtr, Berlin:2017ife, Kim:2019udq}.  
It would be interesting to consider possible mechanisms for setting the thermal relic abundance of ultra massive DM that would naturally lead to accessible signals in direct detection experiments or at accelerators. Such a scenario may be motivated by possible multi-scatter \cite{Bramante:2018qbc, Davoudiasl:2018wxz, Bramante:2018tos,Bramante:2019yss} or single scatter signals in current or near future detectors. 

In this work, we examine whether ultra massive DM particles could be produced initially as a thermal relic, but with a much smaller mass. This would suggest that the thermal relic abundance is set to very small values, which will yield the correct energy density once DM attains a large mass later.  We propose a scenario in which the late time slow-roll of an ultra-light modulus scalar field drives the DM mass to values $\gg 100$~TeV. We note here that similar scenarios were constructed in Refs.~\cite{Hui:1998dc, Zhao:2017wmo}, but with different phenomenology and physics outcomes.  

Intuitively, the small relic abundance implies very large annihilation cross sections.  
We will illustrate through a simple model that this typically points to an initial DM mass $\lsim$~GeV and associated dark states that mediate ``secluded'' annihilations.  
However, once the DM becomes very heavy, it maintains its 
strong coupling to light mediators which would 
naturally lead to detectable signals in the laboratory, such as through direct detection of DM or searches for the mediators in accelerator experiments.  We will refer to our proposed DM candidate 
as a THermal Ultra Massive Particle (THUMP). 

In the following sections, we outline a simple model that realizes the above scenario using a Dirac fermion for DM and a light (sub-GeV) vector particle, which we take to be a ``dark photon'' coupled to the SM through kinetic mixing.  One could invoke a late phase transition (For very recent work in this direction, see Refs.~\cite{Baker:2019ndr, Heurtier:2019beu, Chway:2019kft}) or a very light modulus that effects the mass variation of the DM particle from light (GeV scale) to very heavy.  Here, we will focus on the second possibility, which will typically lead to a long range force acting on THUMPs and could possibly offer further astrophysical signatures. 

\section{A Specific Model}
\begin{figure}[t]
\includegraphics[scale=0.7]{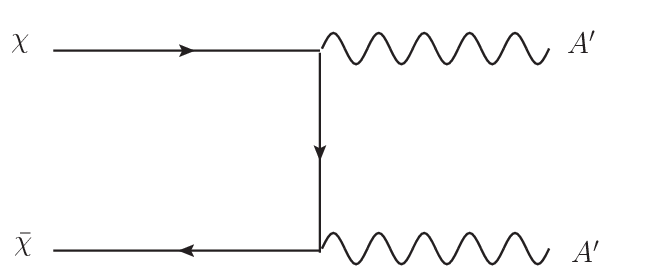}
\caption{Annihilation of DM $\chi$ into ``dark photon'' pairs, which later decay into SM states (only $t$-channel shown).}
\label{chi-anni}
\end{figure}

Let us first consider a DM particle $\chi$ with mass $m_{\chi}^{i}$ in the early Universe, i.e. before Big Bang Nucleosynthesis (BBN). Here, $\chi$ is assumed to have charge $Q_D$ under a dark $\uprime$ interaction mediated by $A_\mu^{\prime}$ of mass $\maprime \lesssim m_{\chi}^{i}$.  We also suppose that 
the DM $\chi$ interacts with a very light scalar $\phi$. The Lagrangian for these interactions is given by 

\bea
-\mathcal{L} &\supset& (\lambda \phi + m_{\chi}^{i}) \bar{\chi} \chi - i g_{D} \,Q_{D} A_{\mu}^{\prime} \bar{\chi}\gamma^{\mu} \chi + \frac{1}{2} \maprime^2 A^{\prime \, 2} \nonumber \\
&+& \frac{1}{2} m_{\phi}^{2} (\phi - \phi_0)^{2}.
\label{eq:lagrangian}
\eea 
In \eq{eq:lagrangian}, $\phi$ sources a large mass $m_{\chi}^{f}\gg m_\chi^i$ for the DM once it starts evolving in its potential. This mass is given by the vacuum expectation value (vev) $\phi = \phi_0$ which yields $m_{\chi}^{f} = m_{\chi}^{i} + \lambda \phi_0$. 
The idea here is that dark matter freeze-out is governed by $m_{\chi}^{i}$ with $\phi$ frozen near $\phi=0$, leading to an efficient annihilation; we will later discuss how this could be arranged due to thermal effects in our scenario. During this time, the thermal relic abundance of DM is set by the annihilation of $\chi$ into pairs of $A^{\prime}$ through a large coupling  $g_D Q_D$ (represented by Fig. \ref{chi-anni}). The vector $A^{\prime}$ then decays into SM particles before BBN; this is thus a secluded annihilation scenario \cite{Pospelov:2007mp}.  The coupling of $A^{\prime}$ to the SM will be assumed to result from kinetic mixing between 
$\uprime$ and SM hypercharge $U(1)_Y$ \cite{Holdom:1985ag}:
\beq
\frac{\eps}{2 \cos \theta_W} F_{\mu\nu}^{\prime} F_Y^{\mu\nu}\,,
\label{kinmix}
\eeq
where $\eps\ll 1$ is the kinetic mixing parameter, $\theta_W$ is the weak mixing angle, and $F_{\mu\nu}$ refers to the associated field strength tensor for each $U(1)$ interaction.
For the rest of this discussion, we set $Q_{D} = 1$.

After DM has frozen out and the relic abundance has been set, the $\phi$ field rolls and oscillates, where its expectation value increases the DM mass to $m_{\chi}^{f}$. Since $ \maprime \lesssim m_{\chi}^{i}$, the annihilation represented by Fig. \ref{chi-anni} is very efficient, especially for $g_D \gsim 1$, resulting in a small relic abundance in the early Universe. The DM energy density at late times is given by $ \rho_{\chi} = m_{\chi}^{f} n_{\chi}$ and the relic abundance is given by 
$\Omega_\chi h^{2} = \rho_{\chi} h^{2}/\rho_{c}$, where $h = 0.678(9)$ is the Hubble expansion rate scale \cite{Patrignani:2016xqp} and $\rho_c \approx 1.05 \times 10^{-5} h^2$ GeV/cm$^3$ is the critical cosmic 
energy density. With the rolling of $\phi$ after freeze out, raising the mass of $\chi$ to 
$m_{\chi}^{f}$, we obtain the DM relic energy density we expect today.  We will next estimate 
the typical parameters that could give rise to the above qualitatively described scenario.

\section{Relic Abundance in the early Universe\label{sec:relic}}

In this work we consider only $\chi$ as the DM and not $\phi$. However in principle $\phi$ could contribute to the DM budget in the Universe, depending on its mass and when it starts oscillating. As indicated in Eq.~\ref{eq:lagrangian}, the potential for $\phi$ is given by $V(\phi) \sim m_{\phi}^{2} (\phi-\phi_0)^{2}$. At early times when $H ~\textgreater ~m_{\phi}$, the $\phi$ field is overdamped and does not oscillate up until $H \lesssim m_{\phi}$.  
Since we are interested in DM number densities that are far smaller than the usual thermal relic values, the freeze out parameter $x_f \gsim 20$. 
Here $x_f \equiv m_\chi/T_f$, where $T_{f}$ is the freeze out temperature.  Given that the large annihilation cross sections required will lead us to $m_\chi\lsim$~GeV, we expect that 
$T_f$ is $\ord{10~\text{MeV}}$, larger than $\sim$~MeV, to avoid disrupting BBN. 
Therefore, a typical expectation in our scenario is that the freeze out Hubble scale $H_f\sim T_f^2/\mP \sim 10^{-14}$~eV, as a rough guide. 

In what follows, we will consider final masses for DM up to $m_\chi^f \sim 10^9$~GeV.  We find that pushing beyond this limit could require dealing with non-perturbative effects that could complicate the analysis and render our estimates unreliable.  However, there should be no obstacle, in principle, to push beyond the above mass.  We note that the final value of the scalar field $\phi_0$ could be much  larger than $\sim 10^9$~GeV, as long as $\lambda \ll 1$.  

Let us consider 
$\phi_0\sim 10^{15}$~GeV, as may be expected in an ultraviolet (UV) framework, corresponding to 
$\lambda \sim 10^{-6}$.  If $m_\phi \lsim 10^{-14}$~eV, then $\phi$ will start to roll {\it after} the freeze out era discussed earlier, corresponding to typical temperatures $\sim 10$~MeV.  Since we are assuming that the initial value of $\phi$ is given by $\phi_i \approx 0$, the energy density initially stored in $\phi$ will be given by $m_\phi^2 \phi_0^2 \lsim 10^{20}$~eV$^4$.  Absent any significant interactions, this energy density would redshift by $(T_{eq}/T_f)^3 \sim 10^{-21}$, where $T_{eq}\sim 1$~eV marks the era of matter-radiation 
equality, with energy density $\ord{T_{eq}^4}$.  Hence, with the above sample values of parameters, it seems that if $\phi$ attains its ground state instantaneously, it does not have enough energy to account for the final energy stored in DM.  Below, we will show that this apparent mismatch 
can be accounted for once the interactions of $\phi$ with the relic $\chi$ particles is properly considered\footnote{We thank Yue Zhao for discussions on this point.}.  We also note that the above initial energy density stored 
in $\phi$ is much less than the radiation energy density of 
$\sim (10 \, \text{MeV})^4$, so the initial potential energy of $\phi$ does not lead to inflation.  

For the above scenario to yield initially small $\chi$ mass, it is required that $\phi_i\approx 0$ 
to begin with.  One can just assume that this is the case, but it would be more interesting if the 
underlying physics could lead to such an initial condition.  Here, we argue that this is indeed the case in our model (see also Ref.~\cite{Hui:1998dc}).  Let us declare upfront that we will ignore possible quantum corrections to the potential for $\phi$.  This question may need to be addressed in a fuller theory describing the above effective 
interactions and could require new dynamics beyond what has been considered here.  While we do not claim a rigorous connection to the Higgs hierarchy problem in the SM, we simply note that the 
seeming insular nature of the SM could be hinting at a more subtle effect for quantum  
contributions to scalar masses.  We do, however, consider the effect of the ambient medium on 
scalar masses.  

For the assumed interactions, the thermal effects of the DM on 
the scalar $\phi$ can be parametrized by $\delta m_\phi^2 \sim \lambda^2 T^2$, \cite{Kolb:1990vq} where 
$T \gsim 10$~MeV and $\lambda\sim 10^{-6}$.  Hence, the mass of $\phi$, considering 
thermal corrections will be given by 
\beq
m_\phi^2(T) \sim m_\phi^2 + \lambda^2 T^2\,,
\label{mphi2}
\eeq
where we have left out $\ord{1}$ factors that would not affect our main conclusions.   
With the above contribution, the temperature dependent value of the scalar vev will be given by 
\beq
\phi(T) \sim \frac{m_\phi^2\, \phi_0}{m_\phi^2 + \lambda^2 T^2}.
\label{phiT}
\eeq
For typical values of parameters, $m_\phi \sim 10^{-14}$~eV, 
$\lambda \sim 10^{-6}$, and $T \sim \ord{10~\text{MeV}}$, we see that 
$\phi(T) \sim 10^{-30} \phi_0$, which is a severe suppression and will ensure that the field is pegged near $\phi = 0$ for $T\gsim m_\chi^i$.

The above formula is valid when the DM is relativistic and in thermal equilibrium with the SM.  However, we would like to keep $\phi \ll \phi_0$ until freeze-out.  Since for the largest values 
of the final DM mass $m_\chi^f \sim 10^9$~GeV considered here we have $m_\chi^i/m_\chi^f \sim 10^{-10}$, we would like 
to have $\phi(T_f)/\phi_0 \lsim m_\chi^i/m_\chi^f$, so that DM mass remains low enough throughout the freeze-out process.  One can easily check that the discussion presented for this choice of parameters can be applied to other large values of $m_\chi^f$ considered in this work. 

We note that $m_\chi^f \sim 10^9$~GeV for the THUMP mass corresponds to a relic number density of $n_\chi\sim 10^{-18} T_f^3$ at freeze out, 
compared to $\sim T^3$ in the relativistic regime.  The in-medium ``tadpole'' term 
$\lambda \phi \, \bar \chi \chi \to \lambda \phi \, n_\chi \sim 10^{-3}\text{~eV}^3 \phi$ would then dominate the vacuum 
term $m_\phi^2 \phi_0 \phi \sim (10^{-4}\text{~eV}^3) \phi$.  The in-medium ``plasma frequency,'' given by $\omega_\phi^2 \sim \lambda^2 n_\chi/m_\chi^i \sim 10^{-17}$~eV$^2$, is much larger than the vacuum parameter $m_\phi^2 \sim 10^{-28}$~eV$^2$.  Given the dominance of the 
medium effects at $T_f\sim 10$~MeV, we note that the value of $\phi$ at this temperature 
is roughly given by $\phi(T_f)/\phi_0 \sim m_\chi^i/(\lambda \phi_0) \sim m_\chi^i/m_\chi^f$, 
which implies that the value of $m_\chi$ is near its initial value all the way down to freeze out temperature and the relic number density calculations are governed by $m_\chi^i$, as assumed in our treatment.  

The above discussion implicitly leads to the conclusion that due to the non-negligible coupling to DM, $\phi$ remains in thermal contact with the SM until temperatures near freeze-out. 
In our analysis the smallest freeze-out temperature is $\sim$10 MeV below which $\phi$ contributes to radiation in a similar way as neutrinos. Below temperatures associated with electron-positron annihilation ($T \sim 1$ MeV), the contribution of $\phi$ to the effective number of neutrino species is given by $\Delta N_{eff} = 4/7 \sim 0.6$. 
For larger freeze-out temperature, this number would be somewhat smaller and hence a typical prediction of our scenario is $\Delta N_{eff} \lsim 0.6$. Interestingly, these values could potentially lead to easing of the tension in the Hubble parameter implied by local and early universe measurements \cite{Bernal:2016gxb, Riess:2016jrr, Aghanim:2018eyx, DEramo:2018vss, Anchordoqui:2019yzc}.

The preceding analysis also implies that the value of $\phi$ remains frozen until the ``tadpole'' term 
becomes dominated by its vacuum value.  For this to happen, $n_\chi$ must be diluted by the 
expansion of the Universe corresponding to a 
temperature $T_*$ at which $(T_*/T_f)^3\sim 1/10$, so that the vacuum tadpole starts to dominate.  We note that this also suggests that the evolution of $\phi$ is not instantaneous and 
the aforementioned mismatch between the initial energy density $m_\phi^2 \phi_0^2$ and 
the eventual DM energy content is remedied.   This is achieved by the dilution of the relic density $n_\chi$ for $T\sim T_*$ at which $\phi$ starts to move away from $\phi(T_f)$.  The above treatment only considers in-medium terms of linear and 
quadratic order in $\phi$.  This truncation is justified since $\phi$ starts out very small due to thermal effects prior to freeze out, as indicated by \eq{phiT}.  Having illustrated that the mass of 
$\chi$ remains close to $m_\chi^i$ with our assumptions about the model parameters, we will next 
examine the annihilation processes that set the relic abundance of $\chi$. 

The DM particle $\chi$ could give the correct relic abundance through the annihilation process given in Fig.\ref{chi-anni} (plus $u$-channel). 
The annihilation cross-section can be expanded in powers of velocity and written as 
\begin{equation}
 \frac{1}{2} \sigma_{\bar{\chi} \chi \rightarrow A^{\prime} A^{\prime}} v \approx a + b v^{2} + \ord{v^{4}}.
 \label{sigv}
\end{equation}

Using the results of Refs.~\cite{Cline:2014dwa, Berlin:2016gtr, Escudero:2017yia}, we find
\begin{eqnarray}
a &=& \frac{g_D^{4} }{8 \pi \, m_{\chi}^{i\,2} } \frac{(1 - r^{2})^{3/2}}{(2 - r^{2} )^{2}}, \nn \\
b &=& \frac{g_D^{4} }{192 \pi \, m_{\chi}^{i\,2} } \frac{\sqrt{1 - r^{2}} (17r^{6} - 36 r^{4} + 28 r^{2} + 24)}{(2 - r^{2})^{4}}.
\label{aandb}
\end{eqnarray}
Here, $r =  \maprime/ m_{\chi}^{i}$, and the $a$ and $b$ coefficients represent the $s$-wave and $p$-wave contributions respectively.
For the values considered in this work, $a \gg b v^{2}$, making this annihilation process $s$-wave dominated.
Expanding \eq{aandb} around $r$ and in the limit that $r \ll 1$, we get
\begin{equation}
\vev{\sigma_{\bar{\chi} \chi \rightarrow A^{\prime} A^{\prime}} v} \approx \frac{g_{D}^{4}}{16\pi \, m_{\chi}^{i\,2}}  \sqrt{1 - r^{2}}.
\end{equation}

Though this is a hidden sector annihilation, the dark sector and SM are in thermal equilibrium in the early Universe. DM $\chi$ freezes out when it is non-relativistic at a temperature 
corresponding to $x = x_{f}$, which is given by 
\begin{equation}
x_{f} =  {\rm ln} \left[\frac{c(c+2)}{4\pi^{3}} \sqrt{\frac{45}{2}} \frac{g_{\chi} }{\sqrt{g_{*}}} ~m_{\chi}^{i} \mP \frac{a + 6 b/x_{f}}{\sqrt{x_{f}}(1 - 3/2 x_{f})} \right].
\label{xf}
\end{equation}
Here we take $c \sim 0.5$, $g_{\chi}$ counts the internal degrees of freedom of $\chi$, which in our case is a Dirac fermion, and hence $g_{\chi} = 4$ \cite{Kolb:1990vq}. The quantity $g_{*}$ counts the relativistic degrees of freedom and is evaluated at freeze-out.
After freeze-out, the relic energy density of DM is given by 
\beq
\Omega_\chi h^2 = \frac{Y_{\infty} s_{0} h^{2}}{\rho_{c}} m_{\chi}^{f}\,,
\label{Omegachi}
\eeq
where $s_{0} = 2891.2 ~\rm cm^{-3}$ is the late time entropy density of the Universe, $\rho_{c}$ is the critical density defined above and $m_{\chi}^{f}$ is the final mass of the DM after $\phi\to 
\phi_0$.
The quantity $Y_{\infty}$ represents the late time comoving number density of DM which is given by
\begin{equation}
Y_{\infty}^{-1} = \sqrt{\frac{\pi}{45}} \sqrt{g_{*}} m_{\chi}^{i} \mP \frac{a + 3b/x_{f}}{x_{f}}.
\label{yinfinity}
\end{equation}
The observed value of the DM energy density is given by $\Omega_{ob} h^2 \approx 0.12$ \cite{Aghanim:2018eyx}.  For any set of input parameters $\{g_D,m_\chi^i, m_{A'}\}$, we can determine the $a$ and $b$ terms in \eq{aandb},   
$x_f$ using \eq{xf}, and $Y_{\infty}$ in \eq{yinfinity}, which together yield $\Omega_\chi h^{2}$ in \eq{Omegachi}. \\

\begin{figure*}[t!]
\includegraphics[scale=0.54]{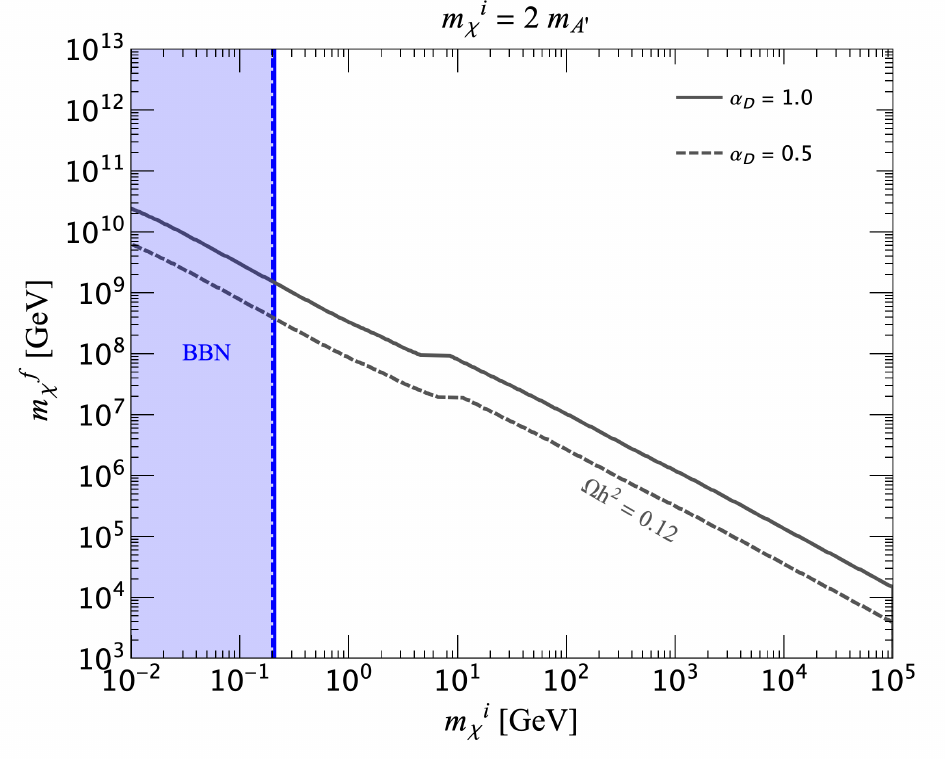}
\hspace*{0.3cm}
\includegraphics[scale=0.54]{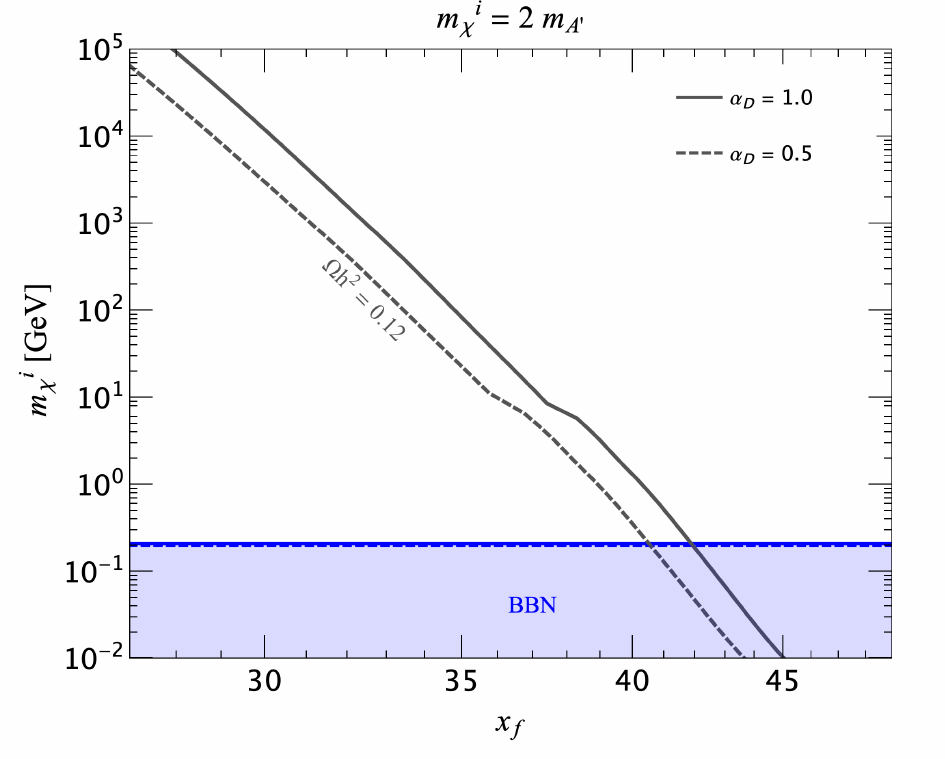}
\caption{Relic abundance in the $m_{\chi}^{f}$ vs $m_{\chi}^{i}$ parameter space on the left and $m_{\chi}^{i}$ vs $x_{f}$ parameter space on the right. The solid diagonal line represents the thermal relic abundance corresponding to $\Omega_{\chi} h^{2} \approx \Omega_{ob} h^{2}$ \cite{Aghanim:2018eyx} assuming $\alpha_{D} = 1.0$ and the dashed line is one with $\alpha_{D} = 0.5$. The blue shaded region represents the BBN bound on this model, in which any initial dark matter mass lower than $\sim 200$ MeV is excluded (see text for more details). }
\label{relic_fig}
\end{figure*}

On the left panel of \fig{relic_fig}, we plot the DM relic abundance in the $m_{\chi}^{f}$ vs $m_{\chi}^{i}$ parameter space and in the $m_{\chi}^{i}$ vs $x_{f}$ parameter space, on the right panel. We assume $m_{\chi}^{i} = 2 m_{A^{\prime}}$ (in general, it is sufficient to assume $m_{\chi}^{i} \gsim m_{A^{\prime}}$), for two values of the dark sector coupling $\alpha_{D} = g_{D}^{2}/4\pi$. The solid (dashed) diagonal line corresponds to the observed relic abundance for $\alpha_{D} = 1.0$ ($\alpha_{D} = 0.5$). 
In this work, we limit the dark sector coupling to $\alpha_{D} = 1.0$ in order to avoid our theory being non-perturbative, as we discussed above. In order to explain the observed relic abundance, we show that the maximum $m_{\chi}^{f}$ we can attain in this model is $\sim 10^{9}$ GeV ($\sim 10^{8}$ GeV) for $\alpha_{D} = 1.0$ ($\alpha_{D} = 0.5$). Furthermore, we note that if we were to choose $\alpha_{D} \sim 4 \pi$ the maximum final mass we would obtain is $\sim 10^{11}$ GeV, though we do not consider this estimate to be reliable. 
In \fig{relic_fig}, we see that near $m_{\chi}^{i} \sim 10$ GeV the relic abundance lines change shape; this is due to the fact $g_{*}(T)$ in our scenario is calculated at freeze-out, and the change in shape corresponds to the QCD phase transition in the early Universe.
DM could annihilate during BBN, thus disrupting elemental abundances during this era. To be conservative, we assume that DM freezes out before BBN.
The shaded blue region shows this BBN bound, which is independent of $m_{\chi}^{f}$, but dependent on $\alpha_{D}$ and $m_{\chi}^{i}$. 
To set this constraint, we assumed that $ T_{\rm BBN} \sim 5$ MeV \cite{Davoudiasl:2010am} and used \eq{xf} to find the initial DM mass $m_{\chi}^{i}$ corresponding to freeze-out during BBN. 
For the benchmark values we have chosen here the boundary occurs near $m_{\chi}^{i} \sim 200$ MeV, as shown by the solid (dashed) line corresponding to $\alpha_{D} = 1.0$ ($\alpha_{D} = 0.5$).
The figure on the right panel can be translated to obtain the temperature at which DM freezes-out for a given $m_{\chi}^{i}$, since $x_{f} = m_{\chi}^{i}/T_{f}$.

\section{constraints and Phenomenology}
In this section we include the various experimental bounds that apply to our model space. In our Galactic neighborhood the virial DM velocity is $v \sim 10^{-3}$. If a DM particle of mass $m_{\chi}^{f}$ scatters with a nucleon of mass $m_{n} \sim 1$ GeV, it will transfer momentum $q \sim m_{n} v \lsim 1$ MeV. 
Throughout this study, we assume $m_{A^{\prime}} \leq ~0.5 ~m_{\chi}^{i}$ and in \fig{relic_fig}, the largest allowed value of $m_{\chi}^{f} \sim 10^{9}$ GeV corresponds to $m_{\chi}^{i} \sim 300$ MeV. Hence for $m_{A^{\prime}} \sim 150$ MeV, $m_{A^{\prime}} \gg q$ and the DM scattering cross-section is given by \cite{Essig:2011nj} 
\begin{equation}
\sigma_{n \chi} \approx \frac{16 \pi \mu_{n\chi}^{2} \varepsilon^{2} \alpha\, \alpha_{D}}{m_{A^{\prime}}^{4}}, 
\label{sig_chin}
\end{equation}
where $\mu_{n\chi}$ is the reduced DM-nucleon mass. For the DM masses considered here, the reduced mass to very good approximation is given by $\mu_{n\chi} \approx m_{n}$.\\

\begin{figure*}[t]
\includegraphics[scale=0.70]{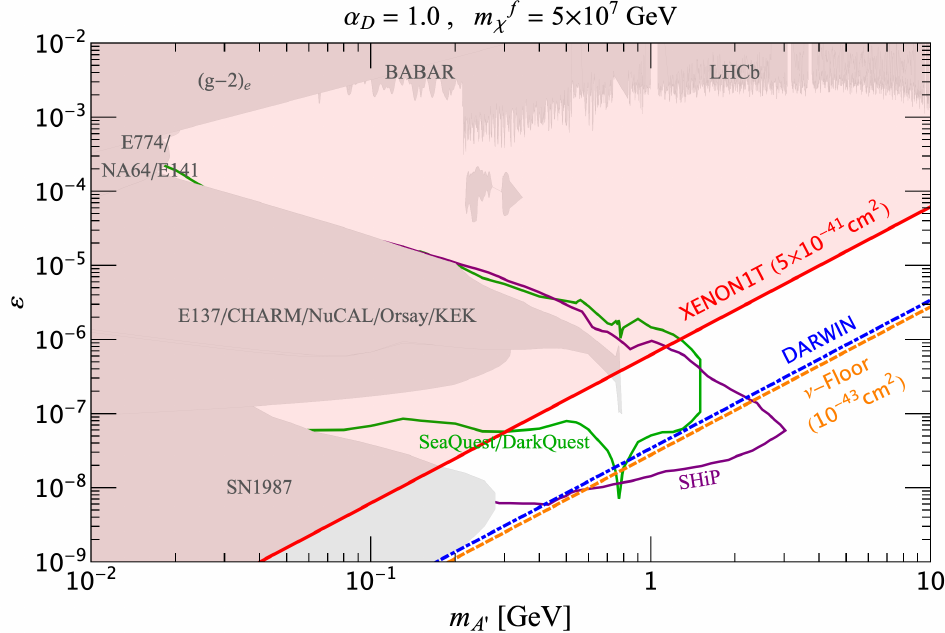}
\caption{Constraints on the $\varepsilon$ vs $m_{A^{\prime}}$ parameter space, taking $\alpha_{D} = 1.0$ and $m_{\chi}^{f} = 5\times10^{7}$ GeV. The grey shaded regions are the current bounds on a visibly decaying dark photon from various accelerator experiments and astrophysical measurements (see text for details). The red shaded region is ruled out by the latest run of the XENON1T experiment. The blue dot-dashed line is the projected sensitivity of the DARWIN experiment and the orange dashed line is the boundary of the neutrino background region for direct detection experiments. The projected sensitivity from the SeaQuest/DarkQuest and SHiP experiments are represented by the regions bounded by the green and purple lines respectively.}
\label{dp_constraints}
\end{figure*}

In \fig{dp_constraints} we show the current model constraints, in the $\varepsilon$ vs $m_{A^{\prime}}$ parameter space, for benchmark $\alpha_{D} = 1.0$ and $m_{\chi}^{f} = 5\times10^{7}$ GeV. 
For the purposes of this study, we consider a mediator mass up to 10 GeV. Given our assumption that $m_{\chi}^{i} = 2 m_{A^{\prime}}$, this corresponds to $m_{\chi}^{f} = 5\times10^{7}$ GeV, which we choose as our benchmark in \fig{dp_constraints}.
The grey shaded areas are current bounds on the visibly decaying dark photon from various fixed target and collider experiments\footnote{For the constraints not depicted here, see Ref.~\cite{Ilten:2018crw}.} \cite{Riordan:1987aw, Bjorken:1988as, Batley:2015lha, Batell:2014mga, Merkel:2014avp, Babusci:2014sta, Anastasi:2016ktq, Banerjee:2016tad, Hanneke:2010au, Lees:2014xha, Aaij:2016qsm, Ilten:2016tkc, Tsai:2019mtm}, and from Supernova 1987A \cite{Chang:2016ntp, Chang:2018rso, Kazanas:2014mca}. 
The most stringent direct-detection bounds on weak scale DM come from the XENON1T experiment and for $m_{\chi}\sim1$ TeV, the limit lies at $\sigma_{\chi n} \sim 10^{-45} \rm cm^{2}$ \cite{Aprile:2018dbl}.
The lack of a DM signal in XENON1T allows us to place a bound on our parameter space, which we obtain by rescaling the XENON1T limit with our mass of $m_{\chi}^{f} = 5\times10^{7}$ GeV. 
This gives us a cross-section upper bound of $\sim 5\times10^{-41} \rm cm^{2}$, which is represented by the red line in \fig{dp_constraints}. The red shaded region is ruled out by the XENON1T experiment. In a similar fashion, the blue dot-dashed line corresponds to $\sigma_{\chi n} \sim 1.5\times10^{-43} \rm cm^{2}$ and is the projected sensitivity of the DARWIN experiment with a 200 ton-year exposure \cite{Aalbers:2016jon}. 
The orange dashed line is the boundary of the neutrino background, below which neutrinos start dominating the nuclear recoil spectrum \cite{Billard:2013qya}.\\

We also show projected sensitivity of future fixed target experiments. The region bounded by the purple solid line is the expected sensitivity reach for SHiP \cite{Alekhin:2015byh,Tsai:2019mtm}, while the green solid line bounds the expected reach for SeaQuest/DarkQuest with $10^{20}$ protons on target \cite{Gardner:2015wea, Berlin:2018pwi, Tsai:2019mtm}. For simplicity, here we only show projections from SeaQuest/DarkQuest and SHiP; for other experiments with similar reach, such as LongQuest or NA62, see Ref.~\cite{Tsai:2019mtm}.
To obtain the bounds shown in grey, we used the publicly available \texttt{DarkCast} code \cite{iltencast} from Ref.~\cite{Ilten:2018crw}. In \fig{dp_constraints}, we show that future experiments such as SHiP or SeaQuest/DarkQuest would be able to probe the $\sim$ GeV-scale mediators of THUMP DM, corresponding to DM-nucleon cross-sections of $10^{-45} - 10^{-40} \rm cm^{2}$. Some of this parameter space may be probed by next generation direct detection experiments, such as DARWIN.
However the GeV-scale mediator signals available to accelerator experiments may provide hints of the dark sector below the neutrino floor.

We note that our model contains a long range force that acts on THUMPs with range typically of the order of $1/m_{\phi} \sim 10^4$ km. This could affect the distribution of DM on macroscopic scales with potentially observable effects. An examination of such effects is beyond the scope of this work. We also point out that a two-loop process mediated by DM and $A'$ couples $\phi$ to ordinary charged matter, however this effect is too suppressed to have any observable effects.

\section{Discussion and Conclusions}
In this work, we proposed a new mechanism for generating the thermal relic abundance of DM with masses above the so-called unitarity limit of $\sim 10^{2}$~TeV. We showed that DM would be composed of GeV-scale WIMPs in the early Universe, before decoupling from the thermal bath.  
If these WIMPs are coupled significantly to other lighter dark sector mediators, they could annihilate efficiently enough that the DM relic abundance after freeze-out is too small, given its initial mass, to constitute the observed DM energy density today.  We postulate that the DM is coupled to an ultra-light scalar $\phi$ which is held near $\phi=0$ in its potential before the DM freezes out.  After freeze-out, $\phi$ starts rolling to some large minimum, sourcing a large mass for the DM, thereby 
raising its energy density to the observed levels.  We call the DM in our scenario a THermal Ultra Massive Particle or ``THUMP.'' 

In a specific model, we assumed that DM, which is initially at GeV scale can couple to a lighter dark photon mediator with strengths of up to $\alpha_{D} = 1.0$.  The strength of this coupling determines how low the DM number density can be after freeze-out, setting the maximum final THUMP mass that would be sourced by the scalar $\phi$ in order to get the correct relic abundance.  

This model provides a connection between ultra heavy DM and GeV-scale mediators. Hence, our proposal motivates DM-nucleon cross-sections that are within reach of the next generation direct detection experiments as well as GeV-scale mediators that can be probed at accelerators. Our scenario takes advantage of the various complementary methods of searching for DM, at both the cosmic and intensity frontiers. We further point out that the model discussed here may be extended to other examples such as a scalar mediator, which would give different phenomenology, however, we leave this for future work.

\textbf{Acknowledgements}
We would like to thank David Curtin, Can Kilic, Michele Papucci, Arvind Rajaraman, Josh Ruderman, Tim Tait, Sean Tulin, Neal Weiner, and Yue Zhao for very helpful discussions. We would also like to thank Yu-Dai Tsai for very helpful correspondence regarding the dark photon bounds and projections.  This work is supported by the United States Department of Energy under Grant Contract DE-SC0012704.

\bibliography{Massive_DM}

\end{document}